\documentclass[5p,twocolumn,authoryear]{elsarticle}

\usepackage{hyperref}
\usepackage{url}

\journal{Speech Communication}

\usepackage{amsmath,amsfonts}
\usepackage{algorithmic}
\usepackage{algorithm}
\usepackage{array}
\usepackage[caption=false,font=normalsize,labelfont=sf,textfont=sf]{subfig} 
\usepackage{textcomp}
\usepackage{stfloats}
\usepackage{url}
\usepackage{verbatim}
\usepackage{graphicx}
\usepackage{booktabs}
\usepackage{comment}
\usepackage{multirow}
\usepackage{colortbl}
\usepackage{xcolor}
\usepackage{bbm}

\definecolor{lightgreen}{RGB}{200, 255, 200}
\definecolor{medgreen}{RGB}{140, 230, 140}
\definecolor{darkgreen}{RGB}{90, 200, 90}

\usepackage{arydshln}

\usepackage{orcidlink}

\begin{document}

\begin{frontmatter}

\title{Crab: Multi Layer Contrastive Supervision to Improve Speech Emotion Recognition Under Both Acted and Natural Speech Conditions}

\author[dca,recod]{Lucas H. Ueda\orcidlink{0000-0002-1029-3420}}
\ead{l156368@dac.unicamp.br}

\author[dca,recod]{João G. T. Lima\orcidlink{0009-0006-6336-8808}}
\ead{j237473@dac.unicamp.br}

\author[dca,recod]{Paula D. P. Costa\orcidlink{0000-0002-1534-5744}\corref{cor1}}
\ead{paulad@unicamp.br}

\cortext[cor1]{Corresponding author.}

\affiliation[dca]{organization={Dept. of Computer Engineering and Automation (DCA), Faculdade de Engenharia Elétrica e de Computação},
            university={Universidade Estadual de Campinas (UNICAMP)},
            city={Campinas},
            country={Brazil}}

\affiliation[recod]{organization={AI Lab., Recod.ai, Institute of Computing},
            university={Universidade Estadual de Campinas (UNICAMP)},
            city={Campinas},
            country={Brazil}}


\begin{abstract}
Speech Emotion Recognition (SER) in real-world scenarios remains challenging due to severe class imbalance and the prevalence of spontaneous, natural speech. While recent approaches leverage self-supervised learning (SSL) representations and multimodal fusion of speech and text, most existing methods apply supervision only at the final classification layer, limiting the discriminative power of intermediate representations. In this work, we propose Crab (Contrastive Representation and Multimodal Aligned Bottleneck), a bimodal Cross-Modal Transformer architecture that integrates speech representations from WavLM and textual representations from RoBERTa, together with a novel \textit{Multi Layer Contrastive Supervision} (MLCS) strategy. MLCS injects multi-positive contrastive learning signals at multiple layers of the network, encouraging emotionally discriminative representations throughout the model without introducing additional parameters at inference time. To further address data imbalance, we adopt weighted cross-entropy during training. We evaluate the proposed approach on three benchmark datasets covering different degrees of emotional naturalness: IEMOCAP, MELD, and MSP-Podcast 2.0. Experimental results demonstrate that Crab consistently outperforms strong unimodal and multimodal baselines across all datasets, with particularly large gains under naturalistic and highly imbalanced conditions. These findings highlight the effectiveness of \textit{Multi Layer Contrastive Supervision} as a general and robust strategy for SER. Official implementation can be found at \url{https://github.com/AI-Unicamp/Crab}.

\end{abstract}


\end{frontmatter}

\section{Introduction}

\noindent Speech emotion recognition (SER) aims to identify the emotional content of speech beyond its semantic meaning~\citep{lech2020real}. The development of effective SER systems has the potential to enhance human-machine interaction significantly~\citep{mustafa2018speech}. Applications range from improved call center services, through the detection of positive and negative emotions in customer speech, to online education scenarios, where instructors can assess students' emotional responses~\citep{madanian2023speech}. 

One of the main challenges in SER lies in identifying and extracting speech representations that are most suitable for computational emotion recognition and discrimination. With recent advances in machine learning, self-supervised learning (SSL) models trained on large volumes of unlabeled data have emerged as powerful tools for speech processing. Models such as HuBERT~\citep{hsu2021hubert}, wav2vec~2.0~\citep{Baevski2020wav2vec2}, and WavLM~\citep{chen2022wavlm} learn informative representations directly from raw audio and are therefore widely adopted as backbone models for downstream tasks. Recent SER approaches leveraging these representations have consistently outperformed traditional handcrafted features, such as filter banks (fbank)~\citep{atmaja2022evaluating}. 

Multimodality has also become a key component in SER systems, often demonstrating superior performance compared to unimodal approaches~\citep{deeb2025enhancing,chu2022self}. In particular, the fusion of audio and textual modalities has been extensively studied and has shown strong performance across multiple datasets. Some works employ SSL models for both speech and text and utilize cross-modal transformers (CMTs) to enable effective interaction between modalities~\citep{khanmemocmt2025}. Other approaches incorporate contrastive learning (CL) to produce more discriminative intermediate representations, leading to improved emotion classification~\citep{khosla2020supervised,wang2023supervised}. 

Despite these advances, most existing studies evaluate their models on well-curated, prompted, and balanced datasets, such as IEMOCAP~\citep{busso2008iemocap} or RAVDESS~\citep{livingstone2018ryeverson}. In practice, however, SER systems should be robust to real-world speech conditions~\citep{koolagudi2012emotion}. Performance on controlled datasets is typically higher than on data containing natural or spontaneous speech~\citep{gomezzaragoza2024emovome,deschamps2021e2e}. Natural emotional speech is inherently imbalanced, which introduces additional challenges for SER. As discussed in~\citep{schuller2011recognising}, datasets containing natural speech are more difficult to collect and model due to this imbalance. 



In this work, we present Crab (Contrastive Representation and Multimodal Aligned Bottleneck) a CMT-based SER model that combines speech representations extracted from WavLM~\citep{chen2022wavlm} with textual representations obtained from RoBERTa~\citep{liu2020roberta}. We further introduce the proposed \textit{Multi Layer Contrastive Supervision} (MLCS), inspired by Contrastive Deep Supervision~\citep{zhang2022contrastive}, in which contrastive objectives are applied at multiple layers of the network to improve representation learning without introducing additional parameters at inference time. To mitigate data imbalance, we employ weighted cross-entropy as the primary classification loss. The proposed approach outperforms baseline methods in terms of Unweighted Average Recall and Weighted Average Recall in both acted datasets (IEMOCAP and MELD~\citep{poria2019meld}) and 

    %

natural condition dataset (MSP Podcast  2.0~\citep{busso2025msppodcastcorpus}). The main contributions of this work are summarized as follows:

\begin{enumerate}
    \item We introduce the \textit{Multi Positive Contrastive Learning} (MPCL) loss as an efficient auxiliary loss to improve emotion classification;
    \item We propose \textbf{Multi Layer Contrastive Supervision (MLCS)}, a training strategy that applies MPCL objectives at multiple stages of a bimodal SER architecture, leading to consistent performance gains on both acted and natural emotional speech conditions;
    \item We conduct an extensive empirical evaluation analyzing the impact of multi layer supervision in comparison to traditional SER training strategies based on single-layer supervision, demonstrating improved robustness to class imbalance and enhanced discrimination of minority emotion classes.
\end{enumerate}

This paper is organized as follows. Section~\ref{sec:related_work} reviews related work on self-supervised representations, multimodal SER, contrastive learning, and deep supervision. Section~\ref{sec:methodology} describes Crab architecture, the MPCL loss function, the MLCS training strategy, and the experimental setup, including datasets and evaluation metrics. Section~\ref{sec:results} presents experimental results and ablation studies that analyze the contribution of multimodality, contrastive supervision, loss design, and different SSL backbones. Finally, Section~\ref{sec:conclusion} concludes the paper.


\section{Related Work}
\label{sec:related_work}

\noindent The use of representations learned by SSL models has been explored in SER for several years and has consistently demonstrated superior performance compared to traditional acoustic features~\citep{atmaja2022evaluating}. Models such as HuBERT~\citep{hsu2021hubert}, wav2vec~2.0~\citep{Baevski2020wav2vec2}, and WavLM~\citep{chen2022wavlm} are trained to extract informative representations from speech signals and are commonly adapted to downstream tasks. In~\citep{atmaja2022evaluating}, a comparative analysis of several SSL models for SER was conducted using a fixed classifier composed of two linear layers with simple average pooling. The authors showed that SSL-based representations significantly outperform traditional filterbank features, attributing this improvement to the ability of SSL models to capture contextual information during pre-training that is absent in handcrafted acoustic features. Among the evaluated models, WavLM Large achieved the best performance. 

Similarly, \citep{kakouros2023speech} evaluated the fine-tuning of these SSL models for emotion classification using different pooling strategies applied to the extracted representations, further demonstrating their effectiveness for SER, with WavLM again achieving the strongest results. \citep{dasaro2024using} also analyzed SSL representations, with a particular focus on wav2vec~2.0, including evaluations in cross-lingual SER scenarios. Beyond standalone SSL features, the integration of additional information sources has also been investigated. In~\citep{deeb2025enhancing}, representations extracted from wav2vec~2.0 were combined with features extracted from mel-frequency cepstral coefficients (MFCCs) by a dedicated layer, illustrating how complementary representations can further benefit SER performance.

\subsection{Multimodality}

\noindent The incorporation of multiple modalities has become increasingly prominent in SER research. In particular, pre-trained text models such as BERT~\citep{devlin2019bert} and RoBERTa~\citep{liu2020roberta} have been integrated alongside speech SSL models to improve emotion classification accuracy. 

A cross-modal transformer (CMT) to fuse speech and text modalities was first proposed in \citep{siriwardhana2020jointly}, where both speech and text SSL models were jointly trained with the CMT architecture. The CMT architecture enables interaction between modalities by using one modality as queries and the other as keys and values, allowing for more effective cross-modal feature integration.

Subsequent works have proposed CMT-based architectures using different encoders and evaluated them on a variety of datasets. \citep{krishna2020multimodal} and \citep{sun2021multimodal} employed bidirectional LSTM-based speech encoders, while \citep{maji2023multimodal} used a bidirectional GRU (bi-GRU) to generate audio representations. These approaches relied on GloVe embeddings for textual input, incorporated CMTs for fusion, and were evaluated on the IEMOCAP dataset. \citep{chu2022self} adopted RoBERTa and HuBERT as modality encoders and introduced an orthogonality regularization term to encourage modality disentanglement, achieving state-of-the-art results on IEMOCAP. More recently, \citep{khanmemocmt2025} proposed MemoCMT, a CMT-based model using HuBERT and BERT, and evaluated it on both IEMOCAP and MELD. Overall, multimodal approaches consistently outperform unimodal systems, underscoring the benefits of incorporating complementary modalities for SER.

\subsection{Contrastive Learning}

\noindent Contrastive learning aims to learn representations by explicitly comparing different samples in the feature space. In practice, neural networks are optimized by contrasting representations of positive and negative pairs, such that representations of positive samples are pulled closer together, while those of negative samples are pushed further apart. This paradigm has been adopted in SER to generate more discriminative representations in conjunction with the standard cross-entropy classification loss between the last layer representation $\mathcal{X}$ and the labels $\mathcal{Y}$, as shown in Equation~\ref{eq:ce_con}.

\begin{equation}
\mathcal{L}_{\mathrm{ser}} =
\mathcal{L}_{\mathrm{CE}}\bigl(m(\mathcal{X}), \mathcal{Y}\bigr)
\;+\;
\alpha \cdot
\underbrace{\mathcal{L}_{\mathrm{CL}}\bigl(m_k(\mathcal{X}), \mathcal{Y}\bigr)}_{\text{contrastive supervision}}
\label{eq:ce_con}
\end{equation}

The use of contrastive learning in SER was first investigated by \citep{lian2018ser}. By combining contrastive loss with cross-entropy loss, their approach achieved improved prediction accuracy by learning more informative representations. Subsequent works extended this idea by applying contrastive learning to align representations from different modalities of the same observation while separating representations from different observations~\citep{weiquan2025coordination,franceschini2022multimodal}. 

Supervised contrastive learning (SCL)~\citep{khosla2020supervised}, which incorporates class label information into the contrastive objective, has also been explored for SER. \citep{wang2023supervised} applied SCL at an intermediate layer of the network, producing representations in which samples from the same class are clustered and samples from different classes are separated. This distance information was subsequently exploited by a KNN classifier at inference time. \citep{alaparthi2023scser} applied SCL to the mean-pooled output of wav2vec~2.0 prior to the classification layer and investigated different data augmentation strategies. \citep{kang2024fcan} introduced an intermediate projection head for SCL, demonstrating effectiveness on both IEMOCAP and MELD. Overall, contrastive learning has proven to be an effective strategy for enriching intermediate representations and improving classification performance in SER.

\subsection{Deep Supervision}

\noindent As neural networks have grown in depth and complexity, training has become increasingly challenging and computationally demanding. Deep supervision, originally proposed for image classification~\citep{wang2015training}, addresses this issue by introducing auxiliary supervision branches at intermediate layers of a network, resulting in additional training objectives alongside the primary loss. This approach was shown to facilitate optimization and improve performance on the ImageNet dataset~\citep{deng2009imagenet}.

Building on the effectiveness of contrastive representation learning, \citep{zhang2022contrastive} proposed contrastive deep supervision, which combines deep supervision with contrastive learning. This method demonstrated superior performance for image classification on both ImageNet and CIFAR-10 without introducing additional parameters at inference time, making it an efficient and computationally inexpensive strategy. The resulting loss function is shown in Equation~\ref{eq:ce_deep_con}.

\begin{equation}
\mathcal{L}_{\mathrm{ser}} =
\underbrace{\mathcal{L}_{\mathrm{CE}}\bigl(m_K(\mathcal{X}), \mathcal{Y}\bigr)}_{\text{standard training}}
\;+\;
\alpha \cdot
\underbrace{\sum_{i=1}^{K-1}
\mathcal{L}_{\mathrm{CL}}\bigl(m_i(\mathcal{X}), \mathcal{Y}\bigr)}_{\text{deep contrastive supervision}}
\label{eq:ce_deep_con}
\end{equation}

To the best of our knowledge, deep supervision has not yet been investigated in the context of speech emotion recognition.

\section{Methodology}
\label{sec:methodology}

\noindent Several recent SER models adopt a bimodal setting by combining SSL-based speech representations with pre-trained language models for textual representation. However, significant challenges remain in scenarios characterized by severe class imbalance among emotions or by the presence of natural speech. We propose an approach based on guiding multiple layers with contrastive learning to assist the model in generating informative intermediate representations for the final classification. The proposed architecture is evaluated on multiple datasets with varying degrees of emotional naturalness, ranging from acted to fully natural speech.

\subsection{Datasets}

\noindent Emotional speech datasets can be categorized as prompted (directly acted) or non-prompted/natural (collected in specific real-world scenarios)~\citep{schuller2011recognising}. In this work, we adopt three datasets that are widely used in the literature and categorize them as prompted, weakly prompted, and natural.\\

\noindent \textbf{IEMOCAP (prompted/acted):} The IEMOCAP (Interactive Emotional Dyadic Motion Capture) dataset is a widely used multimodal corpus for emotion analysis, collected at the University of Southern California. It consists of approximately 12 hours of audiovisual data recorded from ten professional actors (five male--female pairs) engaged in dyadic interactions. The dataset includes both scripted dialogues and spontaneous improvised scenarios designed to elicit emotional expressions. Speech is captured using high-quality microphones and synchronized with video and detailed motion capture data, including facial expressions, head motion, and hand gestures. Utterances are segmented at the speaker-turn level and annotated by multiple human evaluators using categorical emotion labels as well as dimensional attributes. In this work, both improvised and scripted utterances were used, comprising 1{,}103 angry, 1{,}708 neutral, 1{,}084 sad, and 1{,}636 happy utterances. Following common practice in the literature, the happy and excited categories were merged, as these four emotions are the most frequently used for system evaluation. We consider this dataset as prompted due to the fact that the actors knew the goal was to build an emotional dataset.\\

\noindent \textbf{MELD (weakly-prompted/acted):} The Multimodal EmotionLines Dataset (MELD)~\citep{poria2019meld} is a large-scale benchmark dataset for emotion recognition in conversational settings, collected from the television series \emph{Friends}. It contains multi-party dialogues annotated at the utterance level with emotion and sentiment labels, covering seven emotion categories (anger, disgust, fear, joy, sadness, surprise, and neutral) and three sentiment classes (positive, negative, and neutral). MELD provides aligned audio, visual, and textual modalities, enabling research on both multimodal and unimodal emotion recognition. The dataset includes speaker identities, dialogue context, and natural conversational dynamics such as interruptions, co-articulation, and overlapping emotional cues, making it particularly suitable for studying emotion recognition in realistic, context-dependent scenarios. The official training, validation, and test splits provided with the dataset were used in this work. We consider this one as weakly-prompted because although during scenes actors were not induced to produce specific emotions the TV series context already presumes a certain level of prompting contextual emotions.\\

\noindent \textbf{MSP-Podcast 2.0 (natural):} The MSP-Podcast 2.0~\citep{busso2025msppodcastcorpus} dataset is the latest version of the MSP-Podcast (Multimodal Speech Podcast) corpus, which is designed for speech emotion recognition under naturalistic conditions. It consists of spontaneous speech segments extracted from diverse podcast recordings, capturing authentic and unconstrained emotional expressions rather than acted or elicited speech. Each speaking turn is annotated by at least five human annotators using both categorical emotion labels (e.g., anger, sadness, happiness, fear, disgust, contempt, surprise, and neutral) and dimensional emotional attributes, including arousal, valence, and dominance, rated on a seven-point Likert scale. Categorical labels are obtained via plurality voting, while dimensional annotations are computed by averaging annotator ratings to ensure reliability. The dataset also provides human-annotated transcripts and force-aligned audio--text pairs, enabling multimodal modeling approaches. To support robust and fair evaluation, the corpus is partitioned into speaker-independent training, development, and test sets. In particular, three different test sets are provided. Test 1 consists of a test partition where all speakers were already seen in training data, Test 2 consists of completely unseen speakers and Test 3 is completely out-of-distribution with unseen speakers and in a different recording conditions. Although the dataset provides audio transcriptions, we employed the Canary-Qwen model, a speech-augmented language model built on top of Canary~\citep{puvvada2024less, żelasko2025training} and Qwen~\citep{yang2025qwen3technicalreport}, to generate all transcriptions, as textual annotations are only available for the training set and are not provided for the test set. This strategy ensures consistency across all data splits in the proposed modeling pipeline.

\begin{table}[ht]
\centering
\caption{Emotion distribution across the training datasets. Since IEMOCAP does not provide official data partitions, the reported values correspond to the total number of available samples. For MELD, the validation and test splits preserve the same emotion distribution as the training set. MSP-Podcast 2.0 includes two test partitions with similar emotion distributions as training and a third test set that is balanced across emotion categories and is only accessible through an online evaluation platform.}
\label{tab:emotion_distribution_counts}
\begin{tabular}{cccc}
\toprule
\textbf{Dataset} & \textbf{Emotion} & \textbf{Count} & \textbf{Proportion} \\
\midrule
\multirow{4}{*}{\centering IEMOCAP}
 & Angry   & 1103 & 0,20 \\
 & Happy   & 1636 & 0,30 \\
 & Neutral & 1708 & 0,31 \\
 & Sad     & 1084 & 0,19 \\
\midrule
\multirow{7}{*}{\centering MELD}
 & Anger    & 1109 & 0,11 \\
 & Disgust  & 271  & 0,03 \\
 & Fear     & 268  & 0,03 \\
 & Joy      & 1743 & 0,17 \\
 & Neutral  & 4710 & 0,47 \\
 & Sadness  & 683  & 0,07 \\
 & Surprise & 1205 & 0,12 \\
\midrule
\multirow{8}{*}{\centering MSP-Podcast 2.0}
 & Anger     & 22609  & 0,16 \\
 & Contempt & 2765  & 0,02 \\
 & Disgust  & 1324  & 0,01 \\
 & Fear     & 794  & 0,01 \\
 & Happiness    & 37048 & 0,27 \\
 & Neutral  & 51149 & 0,37 \\
 & Sadness      & 18256  & 0,13 \\
 & Surprise & 3220  & 0,03 \\
\bottomrule
\end{tabular}
\end{table}

\subsection{Crab Model}

\begin{figure*}[!t]
    \centering
    \includegraphics[width=1.0\textwidth]{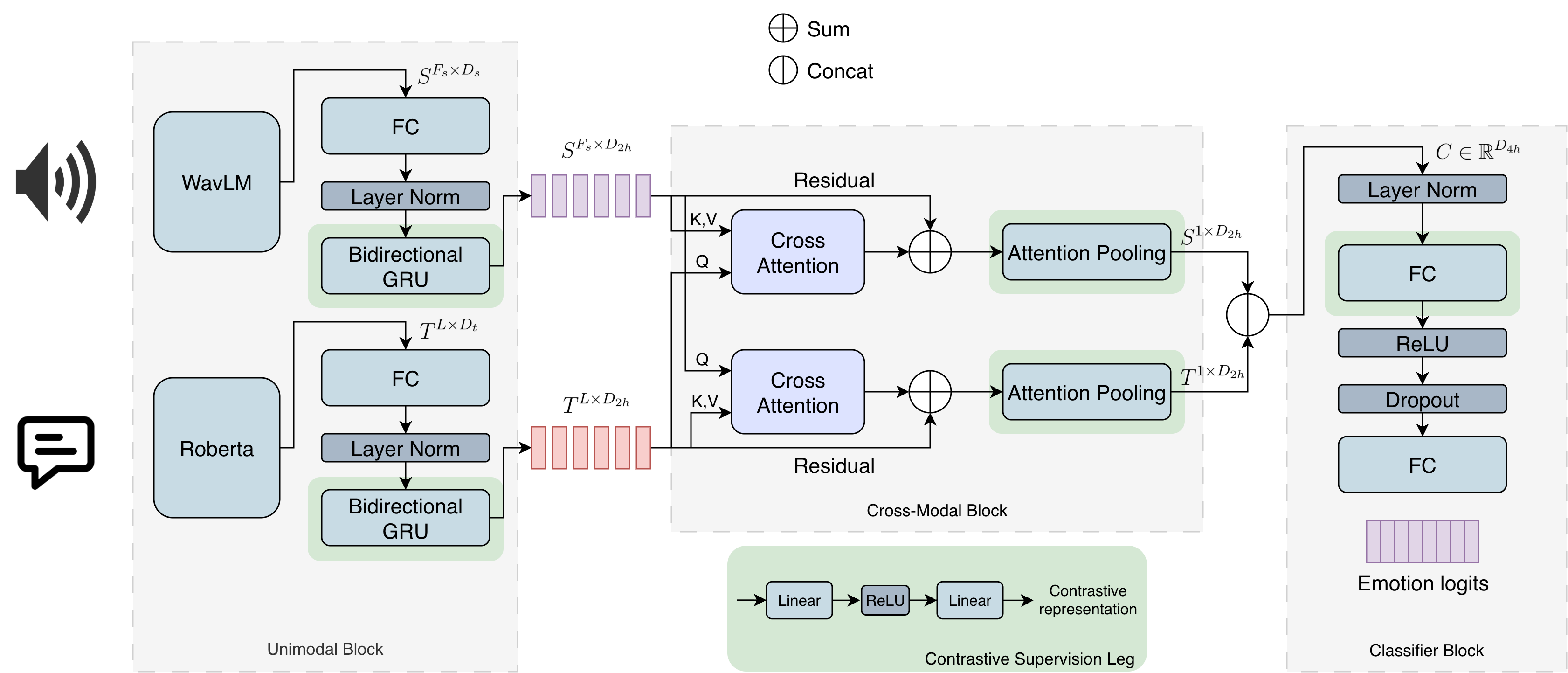}
    \caption{Crab model architecture. The modules highlighted in green in each block have the Contrastive Guidance Leg attached and are optimized using the proposed MPCL contrastive loss.}
    \label{fig:model_arq}
\end{figure*}

\noindent The proposed model is based on a bimodal Cross-Modal Transformer (CMT) architecture that considers both audio and text modalities. For each modality, a pre-trained model is used to extract representations for the proposed architecture. In addition, we introduce a Contrastive Guidance Leg at multiple layers of the model, guiding intermediate representations to become contrastive across different emotions. Figure~\ref{fig:model_arq} illustrates the overall architecture of the proposed model.\\

\subsubsection{Cross-Modal Transformer}

The architecture of the proposed model is a bimodal CMT based on WavLM Large for speech and RoBERTa Large for text. It is divided into three main blocks: the unimodal block, the cross-modal block, and the classifier block. These blocks are highlighted in gray in Figure~\ref{fig:model_arq}, and each is described below.\\

\noindent \textbf{Unimodal block:} Speech features are extracted at the frame level as $S^{F_s \times D_s} \in \mathbb{R}^{F_s \times D_s}$, where $F_s$ is the number of speech frames and $D_s$ the feature dimensionality. Text features are encoded at the token level as $T^{L \times D_t} \in \mathbb{R}^{L \times D_t}$, where $L$ denotes the number of text tokens and $D_t$ the dimensionality of the text features. Both representations are first passed through a fully connected (FC) layer that projects them into a common hidden dimension $h=512$, followed by layer normalization to normalize values across modalities. Each modality is then processed by a bidirectional gated recurrent unit (bi-GRU), producing representations by concatenating the forward and backward hidden states ($S^{F_s \times D_s} \rightarrow S^{F_s \times D_{2h}}$, $T^{L \times D_t} \rightarrow T^{L \times D_{2h}}$).\\

\noindent \textbf{Cross-modal block:} The two modalities are aligned using a cross-modal attention layer based on a multi-head attention mechanism. One modality is used as the query, while the other provides the keys and values, enabling alignment of their representations. The output of the cross-modal attention is added to the corresponding GRU output through a residual connection, balancing modality-specific information with cross-modal interactions. To reduce overfitting, the number of attention heads is set to one. After alignment, features are aggregated using attention pooling, collapsing each modality into a single dense representation weighted by learned attention coefficients ($S^{F_s \times D_{2h}} \rightarrow S^{1 \times D_{2h}}$, $T^{L \times D_{2h}} \rightarrow T^{1 \times D_{2h}}$). The attention pooling operation is defined in Equations~\ref{eq:softmax} and~\ref{eq:attention_pool}:

\begin{equation}
    w_i = \frac{\exp\!\left(\frac{r_i^\top m}{\sqrt{D}}\right)}
    {\sum_{l=1}^{L} \exp\!\left(\frac{r_l^\top m}{\sqrt{D}}\right)},
    \label{eq:softmax}
\end{equation}

\begin{equation}
    r = \sum_{i=1}^{L} w_i r_i,
    \label{eq:attention_pool}
\end{equation}

\noindent where $w_i$ denotes the learned attention weight, $r_i \in \mathbb{R}^D$ represents the $i$-th token or frame feature of dimensionality $D$, and $m \in \mathbb{R}^D$ is a trainable parameter. The pooled representation $r$ corresponds to a weighted average of modality features. All pooled representations are concatenated into a single dense vector $C \in \mathbb{R}^{D_f}$, where $D_f = 4h$.\\

\noindent \textbf{Classifier block:} Emotion logits are obtained by passing $C$ through a classifier composed of a layer normalization, an FC layer, a rectified linear unit (ReLU) activation, and a final FC layer that outputs the emotion-specific logits.\\

\subsubsection{\textit{Multi Layer Contrastive Supervision} (MLCS)}

\noindent Each layer highlighted in green in Figure~\ref{fig:model_arq} is equipped with a Contrastive Supervision Leg (CSL). The CSL receives intermediate representations and encourages them to become contrastive with respect to emotional categories, providing discriminative information throughout the network rather than relying solely on the final classification loss. Each CSL consists of a fully connected (FC) layer followed by a ReLU activation and a second FC layer that projects the representation into a contrastive embedding space. We adopt the multi-positive contrastive learning (MPCL) loss~\citep{tian2023stablerep} as the supervision signal for each CSL.

We define a contrastive categorical distribution \( q \) that measures the relative compatibility between an anchor \( a \) and each candidate \( b_i \):

\begin{equation}
q_i = \frac{\exp\!\left(a \cdot b_i / \tau\right)}
{\sum_{j=1}^{K} \exp\!\left(a \cdot b_j / \tau\right)},
\end{equation}

where \( \tau \in \mathbb{R}_+ \) is a temperature parameter that controls the sharpness of the distribution, and both \( a \) and \( b_i \) are assumed to be \(\ell_2\)-normalized.

The target categorical distribution \( c \) is defined as:

\begin{equation}
c_i =
\frac{\mathbbm{1}_{\text{match}}(a, b_i)}
{\sum_{j=1}^{K} \mathbbm{1}_{\text{match}}(a, b_j)},
\end{equation}

where \( \mathbbm{1}_{\text{match}}(\cdot,\cdot) \) equals one if the anchor and candidate share the same emotional label, and zero otherwise.

The MPCL loss is then defined as the cross-entropy between the target distribution \( c \) and the predicted distribution \( q \):

\begin{equation}
\mathcal{L}_{\mathrm{MPCL}} = - \sum_{i=1}^{K} c_i \log q_i.
\end{equation}

Unlike contrastive deep supervision, which applies supervision to all layers of the architecture, we restrict the CSLs to the proposed bimodal CMT architecture, excluding the standalone WavLM and RoBERTa encoders. For this reason, we refer to our approach as \textit{Multi Layer Contrastive Supervision} (MLCS). In total, CSLs are applied to five layers of the proposed architecture. The first two CSLs are attached to the unimodal blocks of each modality, injecting MPCL-based supervision at the modality-specific level. Since these representations retain temporal dimensions, the CSL inputs are obtained via mean pooling over the bi-GRU outputs. Two additional CSLs are applied after the attention pooling layers, guiding the cross-modal representations to be contrastive with respect to emotional categories. Finally, a CSL is attached to the first linear layer of the classifier, introducing contrastive information into the fused multimodal representation.

The model is optimized using the standard cross-entropy loss combined with the average MLCS loss produced by all CSLs, weighted by a factor~$\alpha$. The final training objective is defined as:

\begin{equation}
\mathcal{L}_{\mathrm{ser}} =
\underbrace{\mathcal{L}_{\mathrm{CE}}\bigl(m_K(\mathcal{X}), \mathcal{Y}\bigr)}_{\text{standard training}}
\;+\;
\alpha \cdot
\underbrace{\frac{1}{|\Omega|}
\sum_{i \in \Omega}
\mathcal{L}_{\mathrm{MPCL}}\bigl(m_i(\mathcal{X}), \mathcal{Y}\bigr)}_{\text{MLCS}}
\label{eq:final_loss}
\end{equation}

\subsubsection{Training Design}

To address class imbalance among emotions, the model is trained using weighted cross-entropy (WCE), where the loss is weighted by the inverse class frequency, following a strategy similar to~\citep{chen2024odyssey}. The class weights are defined as:
\begin{equation}
    w_{prior} = \left\{ w_j = \frac{N}{N_j} \; \middle| \; j \in [1,\ldots,E] \right\},
    \label{eq:wprior}
\end{equation}
where $N$ denotes the total number of samples, $N_j$ the number of samples in class $j$, and $E$ the total number of emotion classes.

The model is trained using the AdamW optimizer with a cosine annealing learning rate schedule. A hierarchical learning rate strategy is employed, where the Crab model parameters are optimized with a learning rate of $1\times10^{-5}$, while the pre-trained encoders (WavLM Large and RoBERTa Large) use a learning rate one order of magnitude lower ($1\times10^{-6}$). All models are trained for 20 epochs with a batch size of 32 and gradient accumulation of 4. All experiments were conducted on a single NVIDIA L40S GPU.

\subsection{Evaluation Metrics}

\noindent In speech emotion recognition (SER), prior work commonly reports the Weighted Average Recall (WAR) and the Unweighted Average Recall (UAR) as primary evaluation metrics. These metrics provide complementary perspectives on model performance, particularly under class imbalance, which is prevalent in naturalistic emotional speech corpora.







WAR can be interpreted as the overall classification accuracy, since it weights each class recall by its empirical class frequency. As a result, WAR is dominated by majority classes and does not adequately reflect performance under class imbalance. In contrast, UAR assigns equal weight to the recall of each class, making it more suitable for evaluating SER systems on naturalistic datasets, where the neutral emotion typically dominates and other emotional states are sparsely represented~\citep{schuller2011recognising,George2024areview}. 

In real-world scenarios, emotions occur with highly non-uniform frequencies, and robustness to class imbalance is therefore a critical requirement for SER models deployed in practical applications. For completeness, we also report the Macro F1-score, which jointly accounts for both precision and recall at the class level. In particular, Macro-F1 is used to evaluate the MSP-Podcast 2.0 dataset, as the official test scoring is conducted through the benchmark platform and relies on this metric. 

\section{Results}
\label{sec:results}

\noindent We first evaluate the proposed model on two benchmark datasets, IEMOCAP and MELD, and compare its performance against four representative baseline systems commonly used in the SER literature.\\

\begin{enumerate}
    \item {\bf{WavLM Baseline}}~\citep{goncalves2024odyssey} is an audio-only SER model built upon the WavLM-large self-supervised representation. A statistics attention pooling layer is applied to the SSL outputs, followed by a classification head to predict emotion logits. This model serves as a unimodal baseline that relies exclusively on acoustic information.

    \item {\bf{FocalSER}}~\citep{chen2024odyssey} is a bimodal SER model that employs cross-attention mechanisms to fuse speech and text representations. The main contribution of FocalSER lies in the use of focal loss to better model minority emotion classes. The original system is an ensemble of multiple models using different SSL encoders for speech and text. In this work, however, we adopt a single-model configuration using WavLM and RoBERTa as SSL encoders, trained with focal loss, to ensure a fair comparison. This approach achieved the best performance in the 2024 edition of the Speech Emotion Recognition in Naturalistic Conditions Challenge.

    \item {\bf{MemoCMT}}~\citep{khanmemocmt2025} is a recent state-of-the-art SER model that employs cross-attention to fuse multimodal representations. It uses HuBERT for speech and BERT~\citep{devlin2019bert} for text as SSL encoders. A distinctive characteristic of MemoCMT is its training strategy with a batch size of one, which eliminates the need for padding or trimming variable-length audio and text sequences. According to the authors, this strategy reduces redundant information and improves robustness to variable-length inputs.

    \item {\bf{Medusa}}~\citep{chatzichristodoulou2025interspeech} achieved the best performance in the 2025 edition of the Speech Emotion Recognition in Naturalistic Conditions Challenge. Instead of cross-attention, Medusa fuses modalities using a deep fusion strategy based on a late-branch architecture. The model is trained with MixUp data augmentation and weighted cross-entropy loss to mitigate class imbalance. Although the original system is an ensemble of multiple models, we evaluate a single-model variant using the proposed weighting strategy and MixUp augmentation. For consistency and fair comparison, we employ WavLM and RoBERTa as the SSL encoders.
\end{enumerate}

\noindent Table~\ref{tab:meld_iemocap} presents a comparative evaluation of all baseline systems and the proposed Crab model. Throughout this work, we follow a consistent visualization scheme: bold values indicate the best performance within each group, underlined values denote the second-best results, and green highlighting is used to emphasize the proposed model or components introduced by our approach.

\begin{table}[ht]
\centering
\caption{Comparison of model performance on IEMOCAP and MELD datasets.}
\label{tab:meld_iemocap}
\begin{tabular}{lcccc}
\toprule
\multirow{2}{*}{\textbf{Model}} & \multicolumn{2}{c}{\textbf{IEMOCAP}} & \multicolumn{2}{c}{\textbf{MELD}} \\
\cmidrule(lr){2-3} \cmidrule(lr){4-5}
 & \textbf{WAR} & \textbf{UAR} & \textbf{WAR} & \textbf{UAR} \\
\midrule
WavLM Baseline & 0.6616 & 0.6832 & 0.3011 & 0.3046 \\
FocalSer                   & 0.6676 & 0.6943 & 0.3751 & 0.4085 \\
MemoCMT                          & \underline{0.6964} & \underline{0.7123} & \textbf{0.6490} & 0.4314 \\
Medusa                          & 0.6605 & 0.6778 & 0.4303 & \underline{0.4548} \\
\rowcolor{medgreen} Crab                  & \textbf{0.7348} & \textbf{0.7585} & \underline{0.6257} & \textbf{0.5565} \\
\bottomrule
\end{tabular}
\end{table}

\noindent The proposed Crab model surpasses all competing methods in terms of UAR, demonstrating its ability to effectively handle both balanced and imbalanced data distributions and to achieve more accurate emotion classification. On the IEMOCAP dataset, where emotion classes are more evenly balanced, the baseline models exhibit broadly comparable performance, with MemoCMT showing a slight advantage over the others. In contrast, on the MELD dataset, where both the training and test splits suffer from substantial class imbalance, MemoCMT achieves the highest WAR among the baselines (0.6490) but underperforms Medusa in terms of UAR, indicating a potential sensitivity to data imbalance. Crab balances WAR and UAR performance on both datasets with the highest performance in terms of UAR and WAR expect by MELD WAR while still being on-par with MemoCMT.

\begin{figure*}[ht]
    \centering

    \subfloat[WavLM Baseline]{%
        \includegraphics[width=0.30\textwidth]{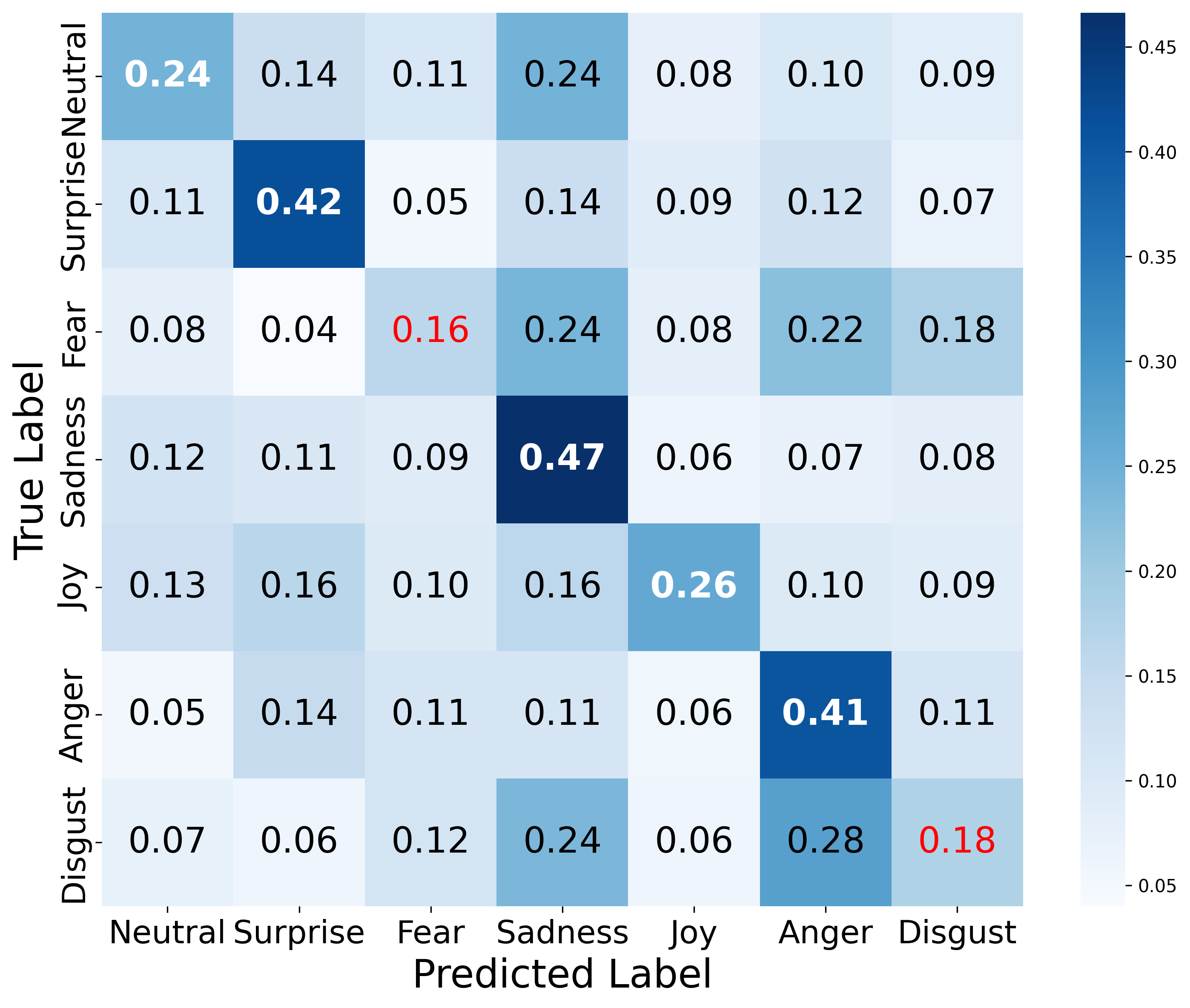}
    }
    \hfill
    \subfloat[FocalSER]{%
        \includegraphics[width=0.30\textwidth]{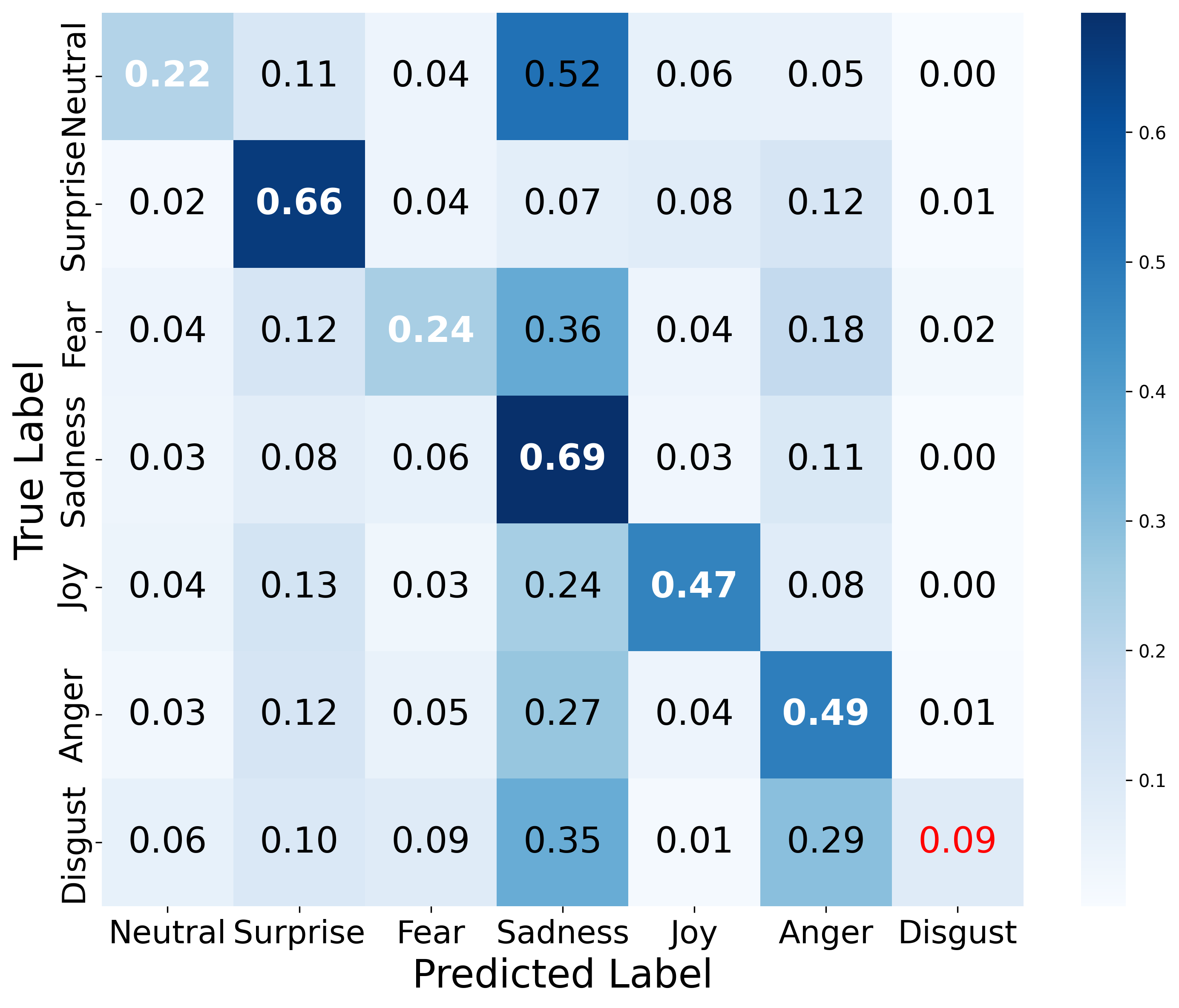}
    }
    \hfill
    \subfloat[MemoCMT]{%
        \includegraphics[width=0.30\textwidth]{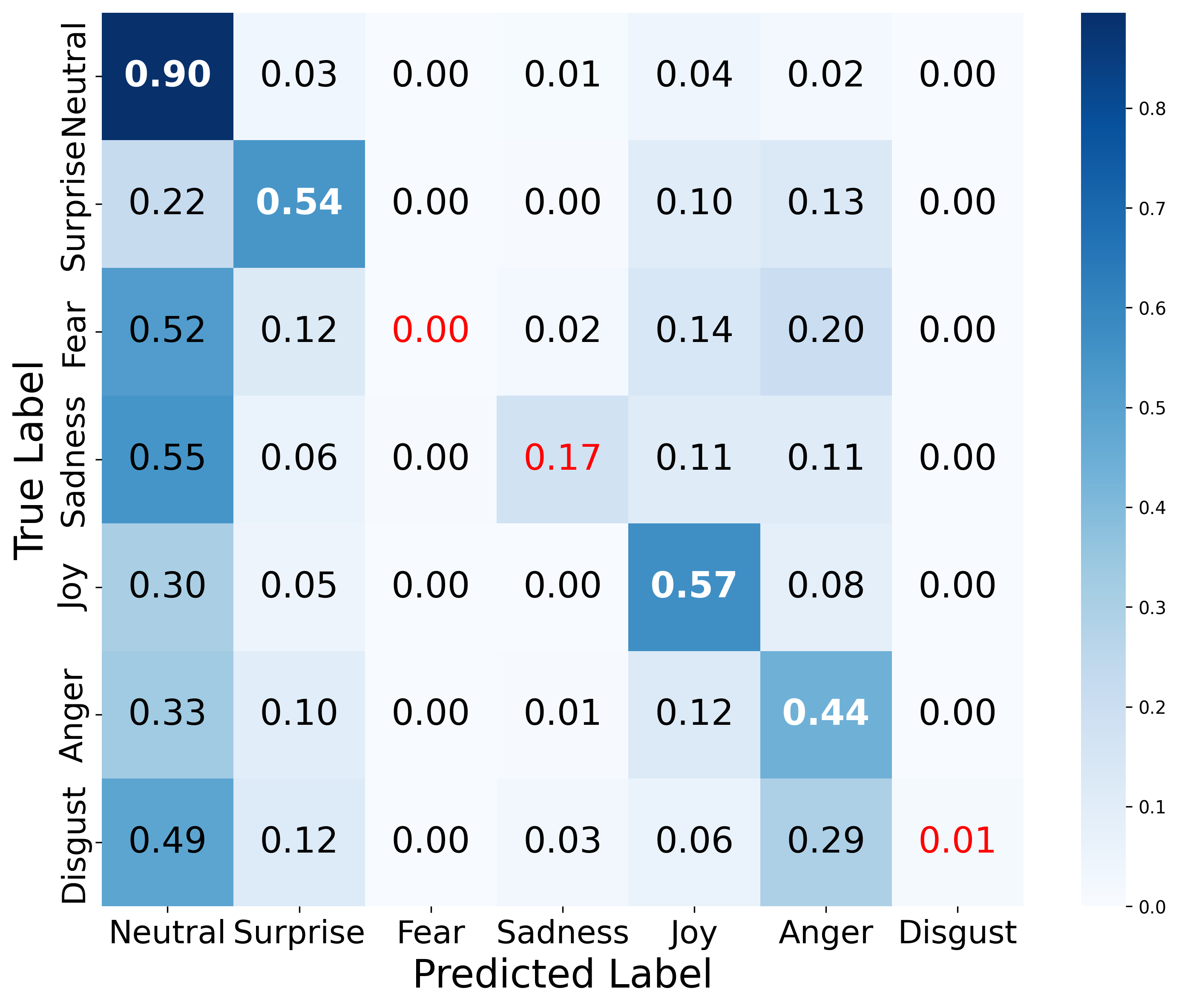}
    }

    \par\smallskip

    \hspace*{\fill}
    \subfloat[Medusa]{%
        \includegraphics[width=0.30\textwidth]{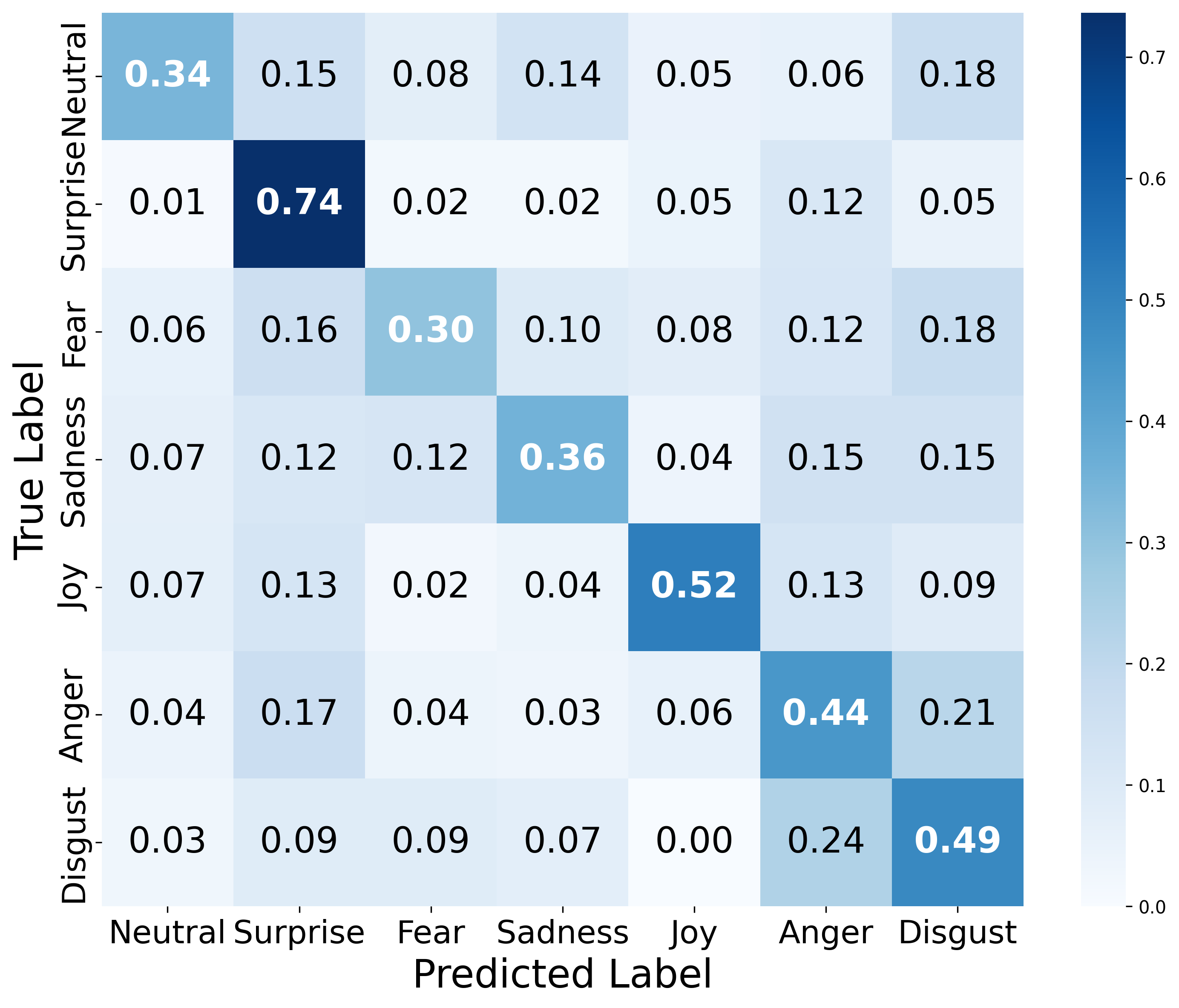}
    }
    \hfill
    \subfloat[Crab]{%
        \includegraphics[width=0.30\textwidth]{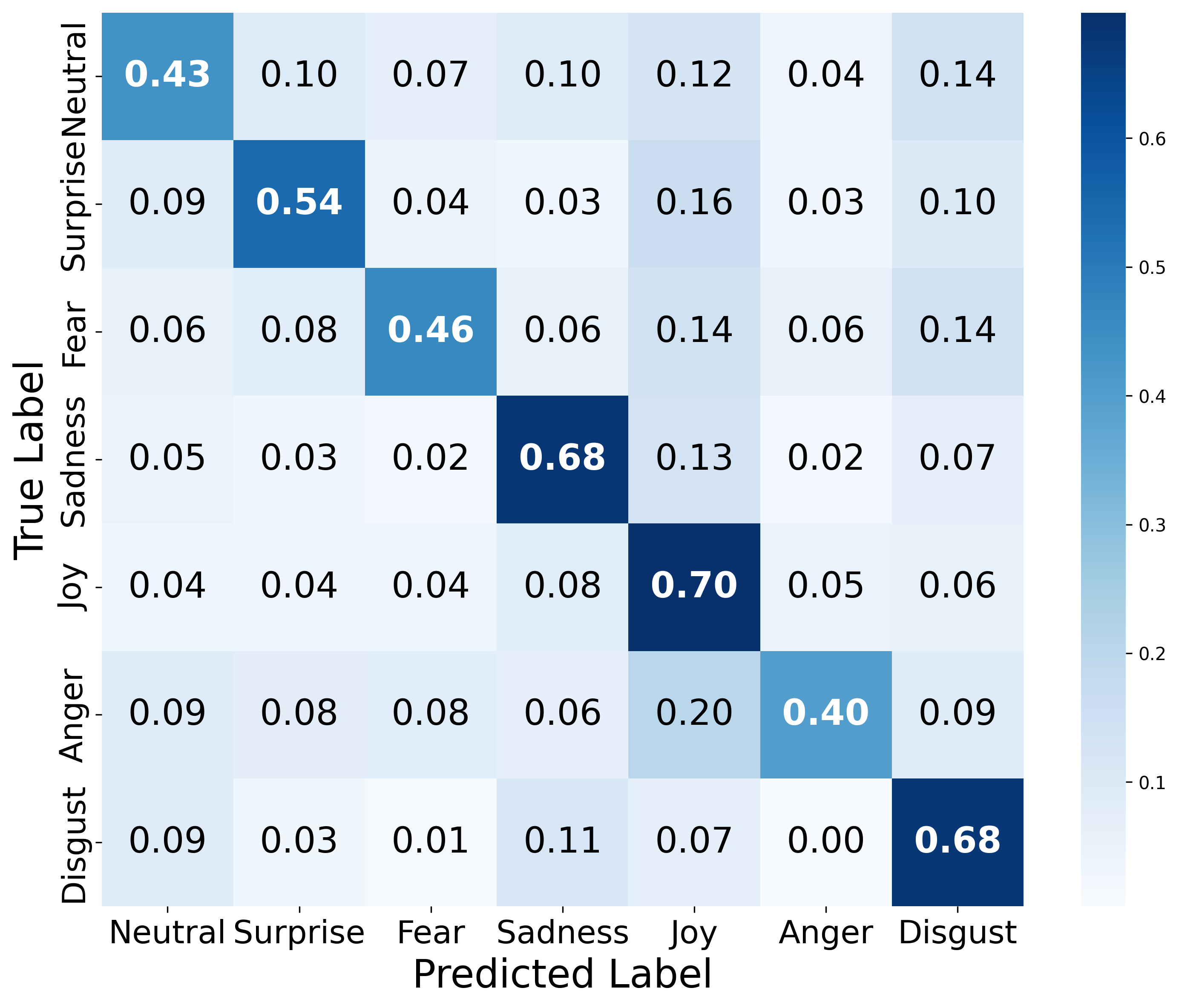}
    }
    \hspace*{\fill}

    \caption{Confusion matrices for MELD dataset across all evaluated models. Values in red on the principal diagonal represent values under a threshold of 0.2. Note that Crab has higher values on the principal diagonal compared to baselines with no emotion category having values under the 0.2 treshold. In particular, the proposed model is robust for the least represented classes (\emph{Fear}, \emph{Sadness}, and \emph{Disgust}).}
    \label{fig:meld_conf_matrices}
\end{figure*}

\noindent It is also worth noting that although FocalSER does not obtain the highest overall metric values, it achieves a higher UAR than the WavLM baseline on the MELD dataset. This result suggests that the use of focal loss benefits the modeling of minority emotion classes compared to standard cross-entropy loss. Furthermore, while MemoCMT remains the second-best performing model in terms of both WAR and UAR, its effectiveness appears to depend on the degree of class balance in the data. This behavior becomes more evident when analyzing the confusion matrices shown in Figure~\ref{fig:meld_conf_matrices}.

The emotions \emph{Fear}, \emph{Sadness}, and \emph{Disgust} are the least represented classes in the dataset, and MemoCMT struggles to model them effectively. In particular, MemoCMT fails to predict the \emph{Fear} emotion entirely and achieves recall values of only 0.17 for \emph{Sadness} and 0.01 for \emph{Disgust}. In contrast, MemoCMT performs best when predicting the \emph{Neutral} emotion, which explains its strong WAR performance. However, the inability to accurately model underrepresented emotions is a critical limitation in real-world applications, where class imbalance is inherent.

The confusion matrices of the WavLM Baseline and FocalSER reveal a bias toward predicting the \emph{Sadness} emotion. Although \emph{Sadness} accounts for only 3\% of the dataset, both models achieve relatively high recall values of 0.47 and 0.69, respectively. Nevertheless, both architectures struggle with the other low-frequency emotion, \emph{Disgust}, which also represents 3\% of the data, achieving recall values of only 0.18 and 0.09, respectively.

Medusa and the proposed model exhibit the most balanced behavior, achieving strong performance while effectively modeling both highly represented and underrepresented emotions. Both models successfully capture the \emph{Neutral} emotion as well as \emph{Fear}, \emph{Sadness}, and \emph{Disgust}. The proposed Crab model consistently achieves higher recall across most emotion categories, as reflected in its superior WAR and UAR compared to Medusa. Specifically, the proposed model outperforms Medusa for all emotions except \emph{Anger}, where its recall is 0.04 lower, and \emph{Surprise}, where it is 0.20 lower. In contrast, it shows notable improvements for \emph{Sadness} (+0.32 recall), \emph{Disgust} (+0.19), and \emph{Joy} (+0.18).

These results demonstrate the effectiveness of the proposed model in achieving strong predictive performance while maintaining the ability to model underrepresented emotions, which is crucial for real-world SER applications. In the following subsections, we further analyze the proposed approach in greater detail.

\subsection{MLCS weight ($\alpha$) in the loss}

\begin{figure}[ht]
    \centering
    \includegraphics[width=\columnwidth]{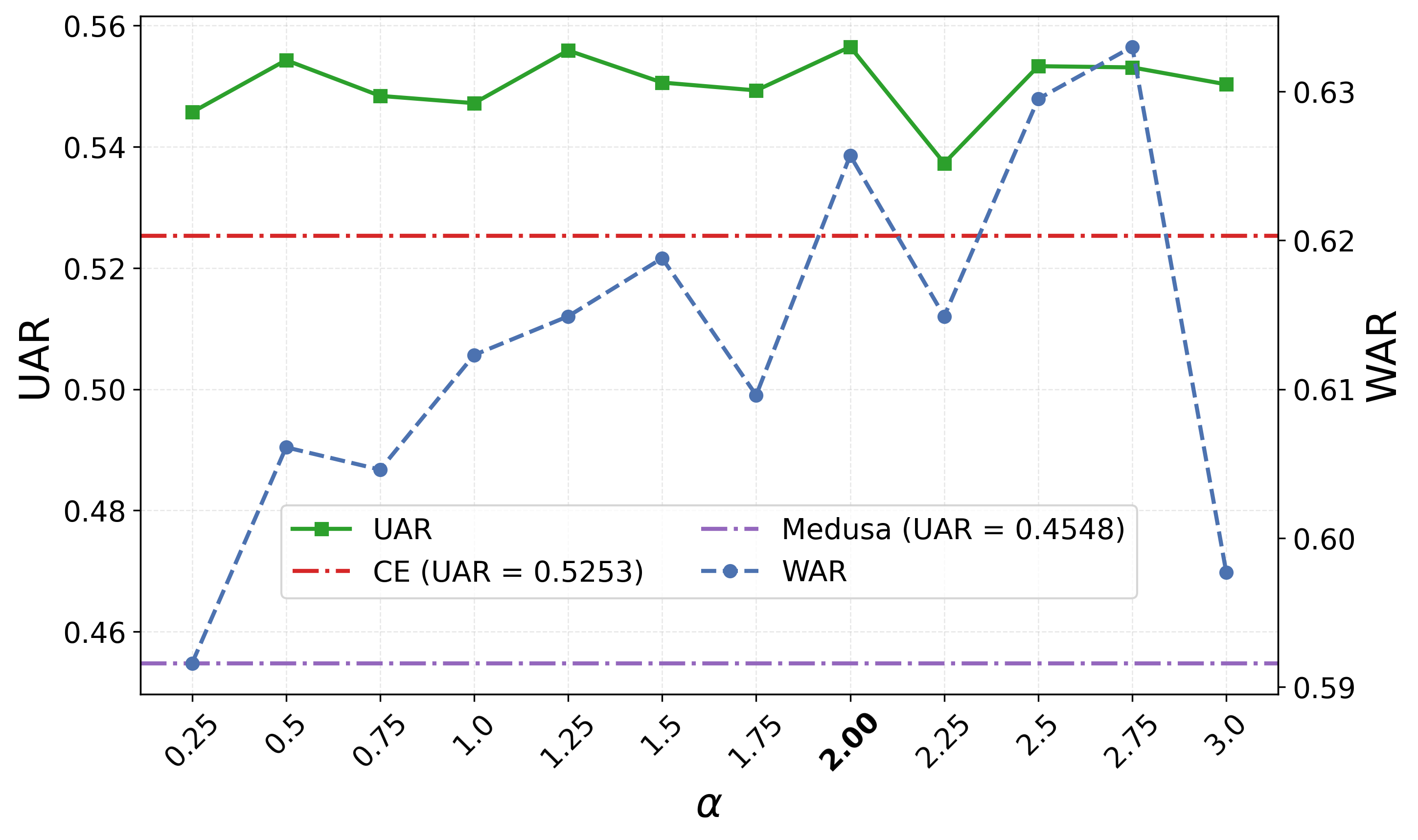}
    \caption{UAR $\times$ WAR performance for different values of $\alpha$. The dashed horizontal lines indicate the proposed model trained only with CE and the Medusa model, using UAR as a reference.}
    \label{fig:uarxwar}
\end{figure}

\noindent To determine an appropriate value for $\alpha$, we evaluated the model performance for values ranging from 0.25 to 3.00. Figure~\ref{fig:uarxwar} shows the UAR and WAR obtained for different values of $\alpha$.

Across the evaluated range of $\alpha$, model performance exhibits limited variation in UAR, peaking at $\alpha = 2.00$, the value adopted in the proposed model. Regarding WAR, a peak value is observed at $\alpha = 2.75$. However, since WAR is influenced by class imbalance, UAR is used as the primary criterion for selecting $\alpha$.

It is worth noting that for all evaluated values of $\alpha$, the model consistently outperforms both the proposed architecture trained only with CE and the best-performing baseline in terms of UAR. This result highlights the robustness and consistency of the performance gains achieved by incorporating MLCS compared to using CE alone, regardless of the weighting applied to the loss.

\subsection{Multimodality}

Many prior works have leveraged both speech and textual information for SER. We investigated the behavior of the proposed model when using each modality individually, compared to the bimodal configuration. To adapt the architecture to a unimodal setting, we remove the cross-modal transformer block shown in Figure~\ref{fig:model_arq}. In this configuration, the pre-trained representations are processed by the bi-GRU, and their outputs are directly aggregated using attention pooling. The CSLs are applied at the same layers as in the proposed bimodal architecture. Table~\ref{tab:ablation_meld_iemocap_modalities} reports the performance obtained for each modality.

\begin{table}[ht]
\centering
\caption{Ablation study of the proposed model on IEMOCAP and MELD datasets.}
\label{tab:ablation_meld_iemocap_modalities}
\begin{tabular}{lcccc}
\toprule
\multirow{2}{*}{\textbf{Model}} & \multicolumn{2}{c}{\textbf{IEMOCAP}} & \multicolumn{2}{c}{\textbf{MELD}} \\
\cmidrule(lr){2-3} \cmidrule(lr){4-5}
 & \textbf{WAR} & \textbf{UAR} & \textbf{WAR} & \textbf{UAR} \\
\midrule
Speech  & 0.6153 & 0.6518 & 0.3724 & 0.3176 \\
Text    & \underline{0.7024} & \underline{0.7198} & \underline{0.6034} & \underline{0.5381} \\
\rowcolor{medgreen}
Speech + Text & \textbf{0.7348} & \textbf{0.7585} & \textbf{0.6257} & \textbf{0.5565} \\
\bottomrule
\end{tabular}
\end{table}

We observe that combining speech and text modalities yields the best performance across all datasets and evaluation metrics. The importance of the text modality is further highlighted by its consistent second-best performance, underscoring the relevance of linguistic content in emotional speech. In particular, on the MELD dataset, the contribution of textual information is more pronounced, with a UAR improvement of 0.2205 compared to using speech alone. This effect may be attributed to the weakly prompted nature of the dataset, as emotions in conversational settings often emerge from contextual cues embedded in the dialogue.

\subsection{Losses}

\noindent To assess the contribution of \textit{Multi Layer Contrastive Supervision}, we evaluate the proposed model under four different training configurations. Specifically, we train the model using only cross-entropy at the output layer (CE), cross-entropy combined with contrastive learning applied solely at the classifier block (CE + MPCL), and multi layer supervision using cross-entropy only (MLS + CE). In addition, we examine the impact of the hierarchical learning rate strategy by training the model with a single learning rate ($1\mathrm{e}{-5}$) for both the main architecture and the pre-trained encoders. Finally, we perform an ablation replacing the MPCL loss with SCL. Table~\ref{tab:training_objectives} summarizes the results of these ablation experiments in comparison to the proposed configuration.

\begin{table}[t]
\centering
\caption{Comparison of training objectives for the proposed model on MELD.
CE stands for Cross-Entropy and MLS denotes Multi Layer Supervision ($\alpha=2$).}
\label{tab:training_objectives}
\setlength{\tabcolsep}{5pt}
\begin{tabular}{lcc}
\toprule
\textbf{Training setting} & \textbf{WAR} & \textbf{UAR} \\
\midrule
CE                              & 0.5797 & 0.5253 \\
CE + MPCL                       & 0.6050 & 0.5375 \\
MLS + CE                         & 0.6077 & \underline{0.5539} \\
MLCS w/ SCL                     & \textbf{0.6337} & 0.5534 \\
MLCS w/o hierarchical lr        & 0.5904 & 0.5182 \\
\rowcolor{medgreen}
MLCS                            & \underline{0.6257} & \textbf{0.5565} \\
\bottomrule
\end{tabular}
\end{table}

The results obtained using CE alone indicate that relying exclusively on standard cross-entropy leads to the weakest performance among all evaluated configurations achieving a UAR=0.5253. Incorporating the MPCL loss alongside CE yields moderate improvements (+0.0122 UAR), demonstrating the benefit of applying contrastive learning to intermediate representations for enhanced emotion classification. Multi layer supervision using CE has a higher impact on performance improving the UAR by 0.0286, highlighting the importance of injecting supervisory signals at multiple stages of the network.

Replacing MPCL with SCL yielded the best WAR results but it still failed to achieve the same proposed model performance in terms of UAR. When the hierarchical learning rate strategy is removed, performance degrades substantially, resulting in the lowest UAR (0.5182). This behavior can be attributed to the fact that the proposed architecture is randomly initialized and using the same learning rate for both newly initialized layers and pre-trained models can propagate noisy gradients into the pre-trained representations, thereby degrading overall performance.

Overall, the proposed training strategy achieves the best results in terms of UAR (0.5565) and the second best in terms of WAR (0.6257), demonstrating that the combination of \textit{Multi Layer Contrastive Supervision} and hierarchical learning rates effectively enhances model performance.

\subsection{Different SSL Models}

\noindent Different speech SSL models can be employed to extract acoustic representations for SER. Among these, three approaches are most commonly used in the literature: wav2vec~2.0, HuBERT, and WavLM.

Wav2Vec~2.0 is based on predicting masked latent speech representations from their surrounding context, following a paradigm analogous to masked language modeling. In addition, it employs contrastive learning to identify the correct discrete latent representation from a set of distractors sampled from other masked regions of the same utterance at different time steps. The self-generated nature of the targets, combined with contrastive learning, encourages the model to prioritize information relevant to linguistic content while deemphasizing factors more closely related to prosody and emotion. This behavior is consistent with its original design goal of automatic speech recognition (ASR)~\citep{chen2022wavlm}.

HuBERT also relies on masked speech prediction. However, instead of self-generated targets, it adopts an iterative re-clustering and re-training strategy. In the first iteration, targets are obtained by clustering MFCC features using $k$-means. In subsequent iterations, targets are generated by clustering the latent representations produced by the previously trained model. Although HuBERT is also primarily optimized for ASR, it has demonstrated stronger performance than wav2vec~2.0 on several downstream tasks.

WavLM further extends masked speech prediction by explicitly incorporating noisy speech denoising, thereby introducing non-ASR-related knowledge during SSL training. Specifically, some input signals are artificially corrupted or overlapped with noise, and the model is trained to predict pseudo-labels corresponding to the original clean speech. Moreover, these SSL approaches rely on large-scale unlabeled corpora for pretraining. Both wav2vec~2.0 and HuBERT are trained on Libri-Light~\citep{kahn2020libri}. However, the predominance of audiobook-style recordings in this corpus introduces a mismatch with real-world data, which can negatively affect performance when adapting the models to more diverse acoustic conditions~\citep{chen2022wavlm}. To address this limitation, WavLM augments its training data with additional corpora such as GigaSpeech~\citep{chen2021gigaspeech} and VoxPopuli~\citep{wang2021voxpopuli}, resulting in more diverse and robust representations. As reported in prior work, WavLM consistently outperforms wav2vec~2.0 and HuBERT on emotion recognition tasks.

We evaluate the proposed model by replacing the speech SSL backbone with each of these alternatives. For each SSL model, we further compare training with standard CE loss against the proposed MLCS strategy, as reported in Table~\ref{tab:ssl_training_meld}. For a fair comparison, only the Large variants of all SSL models are considered.

\begin{table}[t]
\centering
\caption{Performance of Crab using different SSL on MELD.}
\label{tab:ssl_training_meld}
\setlength{\tabcolsep}{5pt}
\begin{tabular}{l ccc}
\toprule
\textbf{SSL} & \textbf{Loss} & \textbf{WAR} & \textbf{UAR} \\
\midrule
\multirow{2}{*}{wav2vec~2.0}
& CE & 0.6023 & 0.5195 \\
& MLCS & 0.6123 & 0.5440 \\
\midrule
\multirow{2}{*}{HuBERT}
& CE & 0.5851 & 0.5229 \\
& MLCS & \underline{0.6230} & \underline{0.5453} \\
\midrule
\multirow{2}{*}{WavLM}
& CE & 0.5797 & 0.5253 \\
& \cellcolor{medgreen}MLCS & \cellcolor{medgreen}\textbf{0.6257}
     & \cellcolor{medgreen}\textbf{0.5565} \\
\bottomrule
\end{tabular}
\end{table}

WavLM Large outperforms the other SSL models (UAR = 0.5565), with HuBERT Large (UAR = 0.5453) achieving the second-best performance, which is consistent with the results reported on the SUPERB benchmark for emotion recognition~\citep{yang2021superb}. Interestingly, regardless of the SSL backbone employed, the use of MLCS consistently outperforms standard CE in terms of both WAR and UAR. This behavior highlights the robustness of the proposed MLCS approach across different types of speech representations.

To further assess the impact of pretraining data and model size, we also conduct an ablation study in which the Crab model is trained using different variants of the evaluated SSL models. Table~\ref{tab:ssl_sizes_training_meld} reports the results for each SSL version, together with the corresponding model size and the amount of data used during pretraining.


\begin{table}[t]
\centering
\caption{Ablation of different versions of wav2vec~2.0, Hubert and WavLM in Crab model.}
\label{tab:ssl_sizes_training_meld}
\begin{tabular}{l l ccc}
\toprule
\textbf{SSL model} & \textbf{Version} & \textbf{Data} & \textbf{WAR} & \textbf{UAR} \\
\midrule
\multirow{3}{*}{wav2vec~2.0}
& Base & 960 hr & 0.6115 & 0.5529 \\
& Large & 960 hr & 0.6123 & 0.5440 \\
& Large LV & 60k hr & 0.6092 & 0.5292 \\
\midrule
\multirow{2}{*}{HuBERT}
& Base & 960 hr & 0.6126 & 0.5443 \\
& Large& 60k hr & \underline{0.6230} & 0.5453 \\
\midrule
\multirow{3}{*}{WavLM}
& Base & 960 hr & 0.6023 & 0.5470 \\
& Base Plus & 94k hr & 0.6123 & \underline{0.5509} \\
& \cellcolor{medgreen}Large & \cellcolor{medgreen}94k hr & \cellcolor{medgreen}\textbf{0.6257}
     & \cellcolor{medgreen}\textbf{0.5565} \\
\bottomrule
\end{tabular}
\end{table}

There exists a positive monotonic relationship for HuBERT and WavLM, whereby larger models trained on more extensive datasets achieve superior performance. In the case of WavLM, the Large model variant, which comprises 316.62M parameters and is trained on 94k hours of data, achieves the highest performance (UAR = 0.5565), followed by the Base Plus variant with 94.70M parameters trained on the same dataset (UAR = 0.5509), and finally the Base model trained on only 960 hours of speech (UAR = 0.5470). These results suggest that the effectiveness of the SSL backbone depends jointly on model capacity and pretraining data scale.

In contrast, wav2vec~2.0 exhibits a negative monotonic relationship. The largest model (Large LV), with 317.38M parameters and trained on 60k hours of speech, achieves the lowest UAR (0.5292), whereas the Base model, with 95.04M parameters trained on 960 hours, attains the highest UAR (0.5529). One possible explanation is that wav2vec~2.0 learns representations that are more tightly coupled to the pretraining data distribution than those of other SSL models due to its contrastive objective. As the amount of pretraining data increases, the negative sampling process may exert a stronger influence on the learned representations. Furthermore, the stronger emphasis of wav2vec~2.0 on ASR-oriented objectives may limit its ability to fully leverage the multimodal capacity of the proposed architecture, resulting in representations that are more aligned with textual content than with emotional cues. This behavior ultimately leads to the lowest performance among the evaluated SSL backbones.

\subsection{Robustness in Naturalistic Conditions}

\noindent In real-world scenarios, emotional expressions are considerably more challenging to model than in prompted or weakly prompted datasets. The pronounced class imbalance that naturally arises in everyday speech poses significant challenges for both training and inference. To evaluate robustness under naturalistic conditions, we conduct experiments on the MSP-Podcast 2.0 corpus and assess each model using data collected in real-life settings. We first evaluated the results on Test 1 and Test 2 sets, which evaluates, respectively, seen and unseen training speakers. The results are presented in Table~\ref{tab:msptest12}

\begin{table}[ht]
\centering
\caption{Comparison of model performance on Test 1 and Test 2 partitions of MSP-Podcast 2.0.}
\label{tab:msptest12}
\begin{tabular}{lcccc}
\toprule
\multirow{2}{*}{\textbf{Model}} & \multicolumn{2}{c}{\textbf{Test 1}} & \multicolumn{2}{c}{\textbf{Test 2}} \\
\cmidrule(lr){2-3} \cmidrule(lr){4-5}
 & \textbf{WAR} & \textbf{UAR} & \textbf{WAR} & \textbf{UAR} \\
\midrule
WavLM Baseline & 0.5131 & \underline{0.4105} & 0.4830 & \underline{0.3131} \\
FocalSer                   & 0.2892 & 0.3360 & 0.1981 & 0.2819 \\
MemoCMT                          & \textbf{0.6239} & 0.3285 & \textbf{0.6058} & 0.2621 \\
Medusa                          & 0.4157 & 0.3549 & 0.4226 & 0.2772 \\
\rowcolor{medgreen} Crab                  & \underline{0.5232} & \textbf{0.4390} & \underline{0.4861} & \textbf{0.3382} \\
\bottomrule
\end{tabular}
\end{table}

The proposed Crab model outperforms all baselines in terms of UAR for both test partitions. For WAR our model is the second best while MemoCMT achieves the highest scores which is aligned with the unbalancing of these partitions where neutral emotion is the most frequent one. This neutral tendency of MemoCMT is also reflected in UAR with this model having the worst performance among all evaluated models. It is worth noting that WavLM Baseline ends up being the second best in terms of UAR indicating that fine tuning the SSL have a higher impact than using pre trained representations. This aligns with our observation that a carefully designed hierarchical learning rate configuration has a significant impact on SER performance. 

In general performance of all models drops from Test 1 to Test 2 partitions since inferring emotions from unseen speakers is a more challenging problem than seen speakers. Despite that proposed Crab model still outperforms other baselines with an UAR of 0.3382.

To assess the model's ability to generalize well to out-of-distribution and unseen speakers we also submitted our model results to the official platform\footnote{\url{https://lab-msp.com/MSP-Podcast_Competition/SERB/}}. Results are presented in Table~\ref{tab:naturalistic_conditions} through accuracy and F1-macro score as metrics.

\begin{table}[ht]
\centering
\caption{Performance of different models under naturalistic conditions on the MSP-Podcast 2.0 Test 3 set. Models with (*) have the results extracted from platform and were not reproduced.}
\label{tab:naturalistic_conditions}
\begin{tabular}{lccc}
\toprule
\textbf{Model} & \textbf{Ensemble} & \textbf{WAR} & \textbf{Macro-F1} \\
\midrule
WavLM Baseline  & No  & 0.3556 & 0.3293 \\
FocalSER        & No  & 0.3272 & 0.3148 \\
MemoCMT         & No  & 0.3184 & 0.2691 \\
Medusa          & No  & 0.3391 & 0.3210 \\
StackingSER (*)     & Yes & 0.4128 & 0.4094 \\
MATER (*)           & Yes & 0.4101 & 0.4097 \\
ABHINAYA (*)          & Yes & \underline{0.4181} & \underline{0.4269} \\
\hline
\rowcolor{medgreen}
Crab            & No  & \textbf{0.4212} & \textbf{0.4313} \\
\bottomrule
\end{tabular}
\end{table}

We observe that the proposed approach outperforms all evaluated baselines, achieving a WAR of 0.4212 and a Macro-F1 score of 0.4313, corresponding to an improvement of approximately 0.08 in Macro-F1 over the best non-ensemble method.

In this evaluation, we also report the results of three additional models submitted to the platform, showing that the proposed single-model approach surpasses highly ensembled state-of-the-art systems. StackingSER~\citep{ueda2025interspeech} employs an ensemble of 12 bimodal and trimodal models, leveraging multiple SSL backbones, including Whisper, HuBERT, and WavLM, which are fused via cross-modal attention. A five-fold Random Forest stacking strategy is then applied to aggregate predictions, resulting in a Macro-F1 score of 40.94

We further compare our results with MATER~\citep{jon2025interspeech} (Multi-level Acoustic and Textual Emotion Representation), a hierarchical framework that integrates acoustic and textual features at the word, utterance, and embedding levels to capture fine-grained prosodic variations and semantic nuances. To address annotator inconsistencies, MATER adopts an uncertainty-aware ensemble strategy and achieves a Macro-F1 score of 40.97\%. 

Finally, we compare against ABHINAYA~\citep{dutta2025abhinaya}, a multimodal ensemble system developed for the Interspeech 2025 Naturalistic SER Challenge. ABHINAYA integrates five complementary models spanning speech-only, text-only, and joint speech–text modalities. The system leverages large-scale pre-trained representations, including WavLM for acoustic modeling and SALMONN-based speech large language models, alongside LLaMA-based text classifiers operating in both zero-shot and fine-tuned regimes. To address the severe class imbalance inherent to naturalistic emotion data, ABHINAYA explores specialized loss functions such as weighted focal loss and vector scaling, and aggregates model predictions via majority voting. This ensemble strategy achieves a Macro-F1 score of 41.81\% on the official test set.

Despite the sophisticated ensembling and stacking techniques employed by these competing approaches, the proposed single-model method achieves superior performance.


\section{Conclusion}
\label{sec:conclusion}

In this paper, we introduced Crab, a multimodal speech emotion recognition framework that combines a Cross-Modal Transformer architecture with a novel \textit{Multi Layer Contrastive Supervision} (MLCS) strategy. By integrating speech representations from WavLM and textual representations from RoBERTa, and by injecting \textit{Multi Positive Contrastive Learning} signals at multiple intermediate layers, the proposed approach encourages the formation of emotionally discriminative representations throughout the network.

Extensive experiments on IEMOCAP, MELD, and MSP-Podcast 2.0 demonstrate that Crab consistently outperforms strong unimodal and multimodal baselines across both balanced and highly imbalanced settings. In particular, the proposed model achieves superior performance under naturalistic conditions, where spontaneous speech and severe class imbalance pose significant challenges. Ablation studies further confirm the individual contributions of multimodality, multi layer supervision, hierarchical learning rate, and the choice of SSL backbones, highlighting the robustness and generality of the proposed training strategy.

\section*{Acknowledgments}
This work is partially funded by CAPES – Finance Code 001. It is also supported by FAPESP (BI0S \#2020/09838-0 and Horus \#2023/12865-8). This project was supported by MCTI under Law 8.248/1991, PPI-Softex, published as Cognitive Architecture (Phase 3), DOU 01245.003479/2024-1.

\section*{CRediT authorship contribution statement}
\textbf{Lucas H. Ueda}: Conceptualization, Methodology, Software, Formal analysis, Investigation, Writing original draft, Visualization. \textbf{João G. T. Lima}: Validation, Investigation, Data curation, Writing review \& editing. \textbf{Paula D. P. Costa}: Writing review \& editing, Supervision, Project administration, Funding acquisition.

\section*{Declaration of competing interest}
The authors declare that they have no known competing financial interests or personal relationships that could have appeared to influence the work reported in this paper.

\section*{Data availability}

The datasets used in this work can be accessed or requested via their official project pages. The MELD dataset is publicly available at https://affective-meld.github.io/. The IEMOCAP (https://sail.usc.edu/iemocap/) and MSP-Podcast Corpus 2.0 (https://www.lab-msp.com/MSP/MSP-Podcast.html) repositories are available upon formal request through their corresponding official pages.

\section*{Declaration of generative AI and AI-assisted technologies in the manuscript preparation process}
During the preparation of this work the author(s) used generative AI tools in order to polish language for clarity and assist with LaTeX formatting of results. After using this tool/service, the author(s) reviewed and edited the content as needed and take(s) full responsibility for the content of the published article.

\bibliographystyle{elsarticle-harv}
\bibliography{refs}  

\vfill

\end{document}